\documentclass[twocolumn,natbib,final]{svjour3}         
\usepackage{graphicx}
\journalname{Microgravity Science and Technology}
\begin{document}

\title{Magnetic gravity compensation\thanks{Presented at the ELGRA Symposium, Bonn, 1-4/9/2009.}}

\author{V. S. Nikolayev\and D. Chatain\and D. Beysens\and G. Pichavant}


\institute{V. S. Nikolayev\and D. Chatain\and D. Beysens\and G. Pichavant \at
              ESEME, Service des Basses
Temp\'eratures,\\ CEA-Grenoble/DSM/INAC,\\ 17 rue des Martyrs, 38054,\\Grenoble Cedex 9, France \\
           \and
           V. S. Nikolayev\and D. Beysens \at
              ESEME, PMMH-ESPCI-P6-P7,\\ 10, rue Vauquelin,\\ 75231 Paris Cedex 5, France\\\email{vadim.nikolayev@espci.fr}
}

\date{Received: 21 October 2009 / Accepted: 18 June 2010}

\maketitle

\begin{abstract}
Magnetic gravity compensation in fluids is increasingly popular as a means to achieve low-gravity for
physical and life sciences studies. We explain the basics of the magnetic gravity compensation and analyze
its advantages and drawbacks. The main drawback is the spatial heterogeneity of the residual gravity field.
We discuss its causes. Some new results concerning the heterogeneity estimation and measurement are
presented. A review of the existing experimental installations and works involving the magnetic gravity
compensation is given for both physical and life sciences. \keywords{Magnetic levitation\and diamagnetic\and
microgravity\and life sciences\and physical sciences\and fluid}
\end{abstract}

\section{Introduction}
\label{intro}

The opportunities of space experimentation are rare and their waiting time is often very long. For this reason,
other ways of achieving reduced gravity (or simulated reduced gravity) are often used as a replacement. Drop
tower and parabolic flight experiments provide short time low-gravity conditions, 4-9 s (Bremen drop tower) and
25 s (ESA Zero-g aircraft). For experiments that require several minutes of low gravity, sounding rockets are
available (ESA Maxus program, 13 min). For experiments that require long low-gravity duration as e.g. in life
sciences, simulation devices like random positioning machines or clinostats can be used. However, all those
means are prohibited in some cases because of security considerations. This concerns the flight experiments
with highly flammable fluids like hydrogen and especially oxygen whose study is extremely important as they are
the fuel components for space propulsion engines.

Another means is used more and more often to achieve long-time low gravity conditions: the magnetic gravity
compensation. Comparing to the other approaches, this means has several undeniable advantages.
\begin{itemize}
    \item It is performed in a ground-based facility with no moving parts so that a good security level can be achieved.
    \item The low gravity duration is unlimited.
    \item In principle, no waiting time.
    \item Reasonable cost.
    \item Possibility of controlling gravity levels (such as corresponding to the Moon, Mars etc.).
    \item Possibility of controlling time variation of gravity, which can reproduce the acceleration (or deceleration) of
    space vehicles.
\end{itemize}
However, drawbacks and important limitations do exist. They will be discussed below. Some additional
explanations and definitions need to be given first.

\subsection{Magnetic gravity compensation versus magnetic levitation}

Magnetic gravity compensation means (total or partial) controlled reduction of the gravity force \emph{at
each point of the object}. This definition is not equivalent to that of magnetic levitation. The latter
requires that the object be suspended, which does not necessarily means that the gravity is compensated
\emph{inside} the object when it is rigid. An example of levitation without gravity compensation is a
transparent bowl placed on a superconductive disk. The bowl contains water with a goldfish. The whole system
is levitated. The photo by \cite{goldfish90} shows that the meniscus of the water is flat, which means that
both water and fish still experience the strong gravity. In what follows, the magnetic gravity compensation
inside \emph{fluids} will be considered. The term magnetic levitation will be rather applied to solid
objects.

\subsection{Magnetic field and magnetic forces}

The magnetic field is characterized by two variables, the magnetic field intensity $\vec{H}$ [A/m] and the
magnetic induction (called also magnetic flux density) $\vec{B}$ [T]. In vacuum, they are related to each other
by the expression
\begin{equation}\label{BHv}
    \vec{B}=\mu_0\vec{H},
\end{equation}
where $\mu_0=4\pi\cdot 10^{-7}$ [T$\cdot$m/A] is a constant called vacuum permeability.

The action of the magnetic field $\vec{H}$ on the matter provokes its own magnetic field called
magnetization:
\begin{equation}\label{M}
    \vec{M}=\chi\vec{H},
\end{equation}
where the coefficient of proportionality $\chi$ is the magnetic susceptibility of the matter. The total magnetic
field is equal to the sum of the external and induced fields,
\begin{equation}\label{Bm}
    \vec{B}=\mu_0(\vec{H}+\vec{M})=\mu\mu_0\vec{H},
\end{equation}
where $\mu=1+\chi$ is the magnetic permeability. The  susceptibility defines the magnetic properties of the
matter. When its absolute value is comparable or larger than unity, the matter is strongly magnetic. This is the
case of ferromagnetic ($\chi\gg 1$) or superconductive ($\chi\approx -1$) substances. In what follows we will
consider only weakly magnetic substances ($|\chi|\ll 1$) that can be either diamagnetic ($\chi<0$) or
paramagnetic ($\chi>0$).

It is important to note that for weakly magnetic substances, $\chi\propto\rho$, where $\rho$ is the mass
density. We will introduce the specific magnetic susceptibility,
\begin{equation}\label{al}
\alpha=\chi/\rho,\end{equation} which characterizes such substances.

Since $\mu\approx 1$ with high accuracy for weakly magnetic substances such as air, the magnetic field created
by a given installation in air is equal to that created in vacuum. For this reason, the $\vec{H}$ value is
related to $\vec{B}$ by the universal relation (\ref{BHv}) and $\vec{B}$ is also often called magnetic field.

Most pure fluids (e.g. H$_2$O, H$_2$, N$_2$) and organic substances are diamagnetic. Some fluids (e.g. O$_2$,
NO) are paramagnetic. The magnetic susceptibility of paramagnetic substances varies with temperature; that of
diamagnetic substances is almost independent on temperature.

The magnetic force that acts on the unit volume of a substance is
\begin{equation}\label{Fm}
   \vec{F}_m=\frac{\chi}{2\mu_0}\nabla (\vec{B}^2),
\end{equation}
where $\nabla$ is the vector gradient operator. The gravity force per unit volume is
\begin{equation}\label{Fg}
    \vec{F}_g=\rho\vec{g},
\end{equation}
where $\vec{g}$ is the Earth gravity acceleration. An ideal compensation is achieved when
\begin{equation}\label{ideal}
    \vec{F}_m+\vec{F}_g=0.
\end{equation}
In a cylindrical $r-z$ reference system where the $z$ axis is directed upwards, this expression is equivalent
to two equations,
\begin{eqnarray}
  \frac{\partial(B^2)}{\partial r}&=&0  \label{Br}\\
  \frac{\partial(B^2)}{\partial z}&\equiv&\nabla(B^2)_z=\frac{2\mu_0g}{\alpha}\equiv G,\label{comp}
\end{eqnarray}
where $\alpha$ is defined by (\ref{al}). It means that for ideal compensation, the magnetic field would need
to satisfy the equation $B=\sqrt{c+Gz}$, where $c$ is an arbitrary constant. It has been shown by
\cite{Quettier} that such a solution of the Maxwell equations for magnetic field does not exist so that the
ideal compensation in any finite volume is impossible. In practice, the ideal compensation is achieved in a
single or at most several points.

The stability of levitation is an important issue and is discussed by many authors starting from
\cite{Braunbek39}. For the purposes of the present study, it is important to mention that the levitation of a
drop (or, generally, of a denser phase) in the surrounding gas is stable for diamagnetic fluids and unstable for
paramagnetic fluids. On the contrary, the levitation of a bubble (or, generally, of a less dense phase) inside
the liquid is stable for paramagnetic and unstable for diamagnetic fluids \citep{MST09}.

\subsection{Required magnetic fields}

\begin{figure}
\centering
\includegraphics[width=0.8\columnwidth]{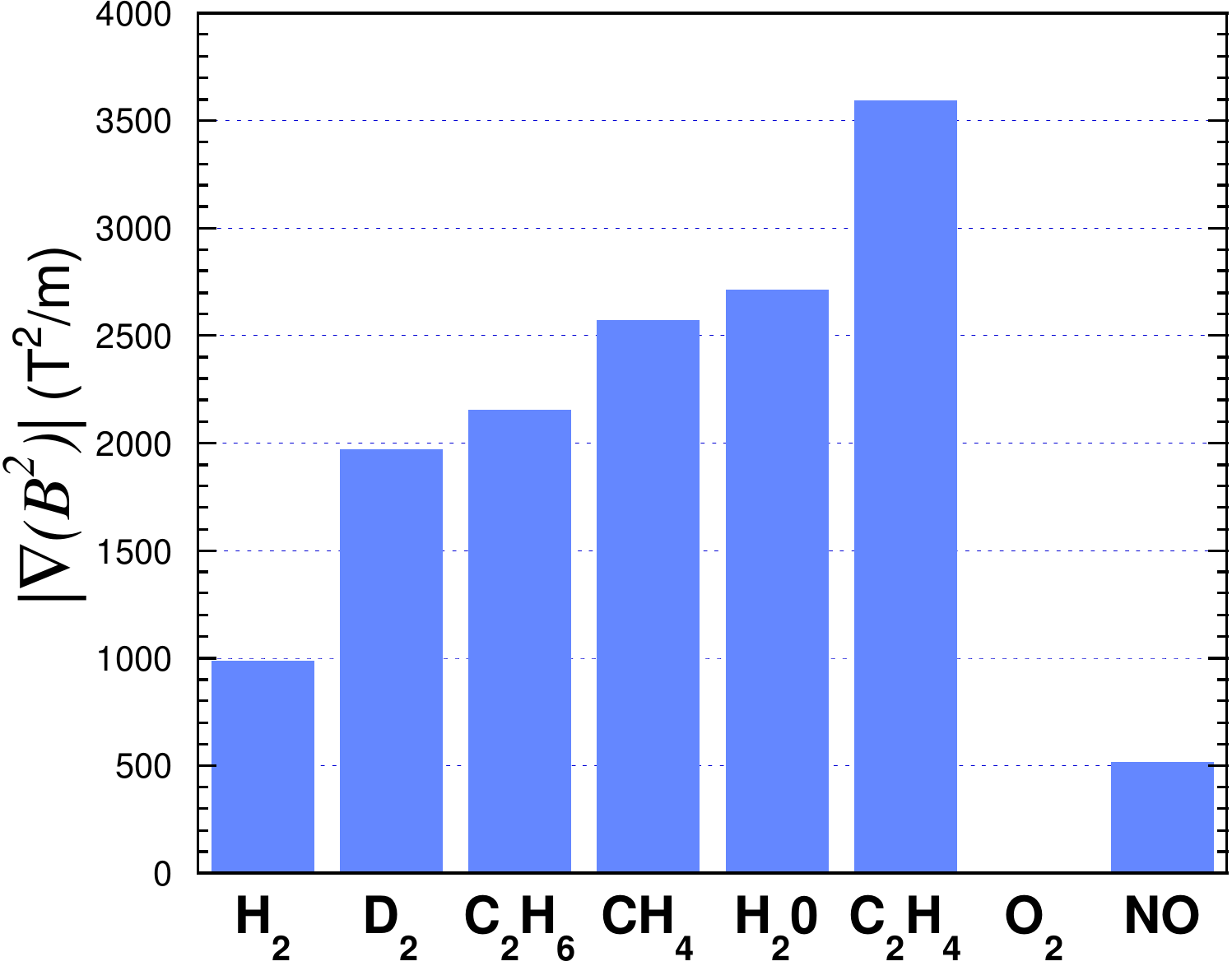}
\caption{The values of $|\nabla(B^2)|$ required for gravity compensation for different fluids. The value for
O$_2$ is about 8 T$^2$/m, which is so small that the corresponding bar is almost invisible. The signs of the
required $\nabla(B^2)_z$ are opposite for paramagnetic (O$_2$ and NO) and diamagnetic (all other) fluids.}
\label{gradB2}
\end{figure}

It is important to underline that the magnetic compensation does not work for all substances at the same
time. The $\nabla(B^2)_z$ value required to compensate the gravity for a particular substance is given by the
material constant $G$ from Eq. \ref{comp}. This value for different substances is shown in Fig.~\ref{gradB2}.
Note that the $G$ value for oxygen is the smallest. In most installations, the magnetic field is created with
one or several co-axial solenoids, for which the radial component of the magnetic force is zero at the axis
so that $|\nabla(B^2)|=|\nabla(B^2)_z|$, where $\nabla(B^2)_z$ can be positive or negative. For this reason,
one speaks often of $|\nabla(B^2)|$ instead of $\nabla(B^2)_z$.
Generally speaking, if a sample is submitted to $\nabla(B^2)_z$ needed to compensate the gravity in a given
substance, the gravity is not compensated for the others.

Note that Eq. \ref{comp} does not involve the density nor the mass of the sample. It means that the gravity will
be compensated independently of the sample mass. If the gravity is compensated for the liquid phase of a
substance, it is also compensated for the gas phase of the same substance, i.e. the buoyancy force for the gas
bubbles or solid crystals in the liquid is compensated either.

In agreement with Eq. \ref{comp}, the ability of a given magnetic installation to compensate the gravity is
characterized by $|\nabla(B^2)|$ that the installation is able to generate.
\begin{figure}
\centering
\includegraphics[width=0.8\columnwidth]{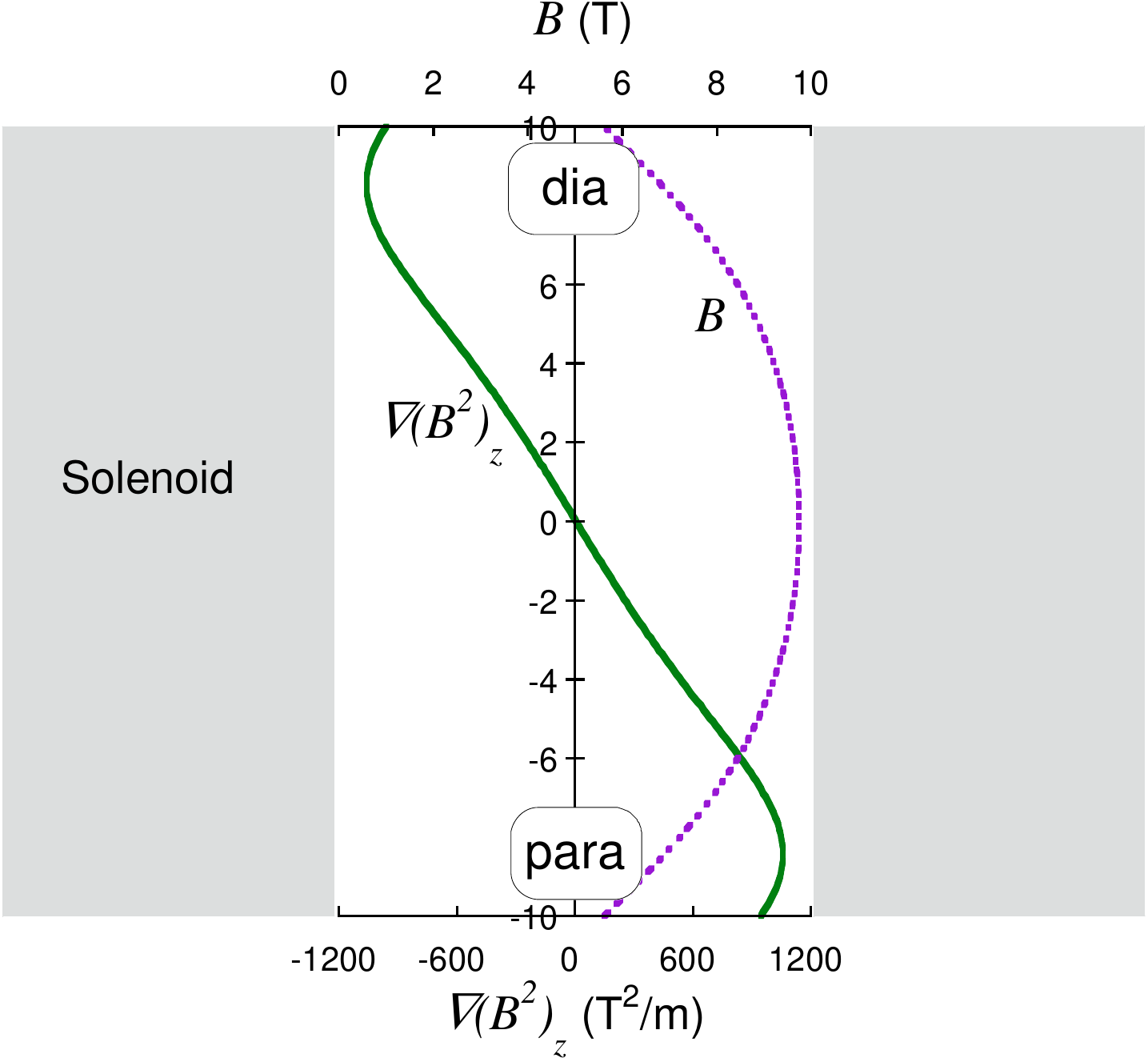}
\caption{An example of the variation of $B$ and $\nabla(B^2)_z$ along the axis $z$ of a solenoid. The
locations appropriate for levitation of dia- and para-magnetic samples are indicated. The HYLDE solenoid
\citep{Cryo02} data are shown. $z$ is measured in cm.} \label{Sol}
\end{figure}
The variation of this value along the axis of a typical solenoid (Fig.~\ref{Sol}) shows that it has two
extrema situated near the ends of the solenoid. These extrema are the most suitable places for the gravity
compensation because they provide the maximum value of $|\nabla(B^2)|$ for a given current in the solenoid.
The upper extremum is a minimum and is suitable for the levitation of diamagnetic substances. The lower
extremum is a maximum and is suitable for paramagnetic substances (Fig.~\ref{Sol}).

\section{Past and present of magnetic gravity compensation}

The bases of magnetic levitation have been put forward by \cite{Braunbek39}. He has succeeded the levitation of
diamagnetic bismuth that has a very low required $|\nabla(B^2)|$. He also provided the theory of levitation.

The first gravity compensation experiments have been realized in the 1960's independently in Berkeley (USA)
by \cite{Lyon65} and in Kharkov (USSR, Ukraine at present) by \cite{Kirichenko68}. They dealt with the
studies of the boiling heat transfer in oxygen that had a very small required $|\nabla(B^2)|$. The magnetic
field was created with resistive solenoids. These studies have been motivated by the importance of oxygen as
a rocket fuel.

The development of superconductive solen\-oids opened the way to their wide use for gravity compensation. It has
been pioneered by \cite{B&T91}, who levitated multiple organic samples, both solid and fluid. The levitation of
the frog embryos by \cite{LevFrogUS} is the first known to us application of the magnetic gravity compensation
in the life sciences. A large number of works on magnetic gravity compensation has been published since then.
\begin{table*}
\centering \caption{Available magnetic gravity compensation installations worldwide} \label{tabI}
\begin{tabular}{lllll}
\hline\noalign{\smallskip}
Location& $B$, T& $|\nabla\,(B^2)|$, T$^2$/m& Bore {\O}, mm& Latest citation \\
\hline\noalign{\smallskip}
Nottingham, UK& 16.5& 2940& 50&\cite{Hill08} \\
Nijmegen, NL& $\sim 17$& $\sim 3000$& 40&\cite{Geim00} \\
Gainesville FL, USA& 15& $ 3000$ & 66& \cite{Mouse09} \\&& 760& 195&\cite{Brooks01}\\
Providence RI, USA& 9.5& $ 3200$ & 11& \cite{Brown04} \\
Xi'an, China&16.12& 3026&51&\cite{China08}\\
Hiroshima, Japan& 15& $\sim 3000$& 50& \cite{Hiroshima} \\
Tohoku, Japan& ~& $\sim 4000$& 52& \cite{Tohoku1} \\
Tsukuba, Japan& 8.5 & 448 & 50&\cite{Tsukuba1} \\& 17& 1600&&\\
Grenoble, France& 10 & 1000 & 50 &\cite{PRL06}\\
& 2& 10& 180& \cite{MST09} \\
\noalign{\smallskip}\hline
\end{tabular}
\end{table*}
The known to us experimental installations available at present for magnetic gravity compensation are presented
in Table \ref{tabI} with their main parameters such as the maximum attainable $|\nabla(B^2)|$ value and the bore
diameter. The latter defines the maximum sample size that can be used. Installations that can attain $\sim 3000$
T$^2$/m may be used for gravity compensation in water or biological tissues that consist mainly of water (cf.
Fig.~\ref{gradB2}); their bore diameters correspond to the thermally insulated part of the bore at room
temperature. Two last lines in the table refer to the installations developed in our group, the HYdrogen
Levitation DEvice (HYLDE) and Oxygen Low Gravity Apparatus (OLGA), respectively.

The physical sciences studies performed with the magnetic gravity compensation in fluids concerned most\-ly
the shape and motion of bubbles and drops. The studies performed at isothermal conditions dealt with the drop
shape \citep{H20SolLev,WunenH2,Hiroshima,Hill08}, drop vibrations \citep{B&T91,HeOscill}, drop coalescence
\citep{B&T91,HeNoncoal}, applications in microfluidics \citep{Lyuksyutov04}, surface instability in the
magnetic field \citep{O2inst03}, and, more recently, a study of the liquid meniscus under fast acceleration
change \citep{GvarELGRA09,OLGA}. The non-isothermal studies concerned boiling
\citep{Lyon65,Kirichenko68,PRL06,MST09}, drop behavior under temperature gradients
\citep{Tohoku1,GradELGRA09}, phase transitions under vibrations
\citep{BeysVib05,BeysVib05a,ActaAstro07,BeysVib08} and gravity influence on flame \citep{Flame08}.

The life sciences studies are concerned with the microgravity influence on protein crystals' growth
\citep{Tsukuba2,China08}, expression of genes \citep{ArabLev07,Coleman2007,ExprELGRA09}, growth of living cells
\citep{Brown04,CellELGRA09,HumELGRA09,Hammer09} or example levitation of small creatures
\citep{LevFrogUS,Mouse09}, and plant morphology \citep{Manzano09}.

\section{Magnetic force heterogeneity issue}

It has already been mentioned that the ideal compensation is achieved only in isolated points. However, it is
possible to approach the ideal compensation conditions within a given accuracy in any volume. The effective
gravity spatial heterogeneity is thus the most important issue that limits the applicability of magnetic gravity
compensation.

The compensation quality can be characterized by the spatial distribution of the effective gravity acceleration
\begin{equation}\label{aeff}
    \vec{a}_{eff}=(\vec{F}_m+\vec{F}_g)/\rho=\vec{g}+\frac{\alpha}{2\mu_0}\nabla (\vec{B}^2)
\end{equation}
defined with (\ref{Fm},\ref{Fg}). In practice, the non-dimensional acceleration heterogeneity
$\vec{\varepsilon}=\vec{a}_{eff}/g$ is more convenient.

There are several possible causes for $\vec{a}_{eff}$ spatial variation. Those related to the spatial
variation of $\nabla ({B}^2)$ manifest itself even in a single-component system, i.e. in a pure fluid sample
where its gas and liquid phases might coexist. Additional spatial variation of the effective gravity appears
in multi-phase samples where $\alpha$ varies. This variation might lead to internal mechanical stresses or
even component displacement in such systems and needs to be analyzed separately for each specific case, for
which the magnetic susceptibility $\chi$ for each of the components needs to be known with precision. Note
that the magnetic susceptibility is well studied for life sciences systems for high frequency magnetic field.
However, $\chi$ value for the \emph{constant} field might be very different and is yet to be determined
experimentally. In the rest of this section we consider only the single-component fluids.

A spatial variation of $\nabla ({B}^2)$ may appear in such fluids because of several reasons. First, there is
a variation of the background force field of the magnetic installation (sec. \ref{Solf}). Second, distortions
can be induced by the sample. These include the variation because of the experimental cell structure (see
below) and the fluid itself. Let us consider each of the heterogeneity causes separately.

\subsection{Background field heterogeneity}\label{Solf}

\begin{figure}
\centering
\includegraphics[width=\columnwidth, clip]{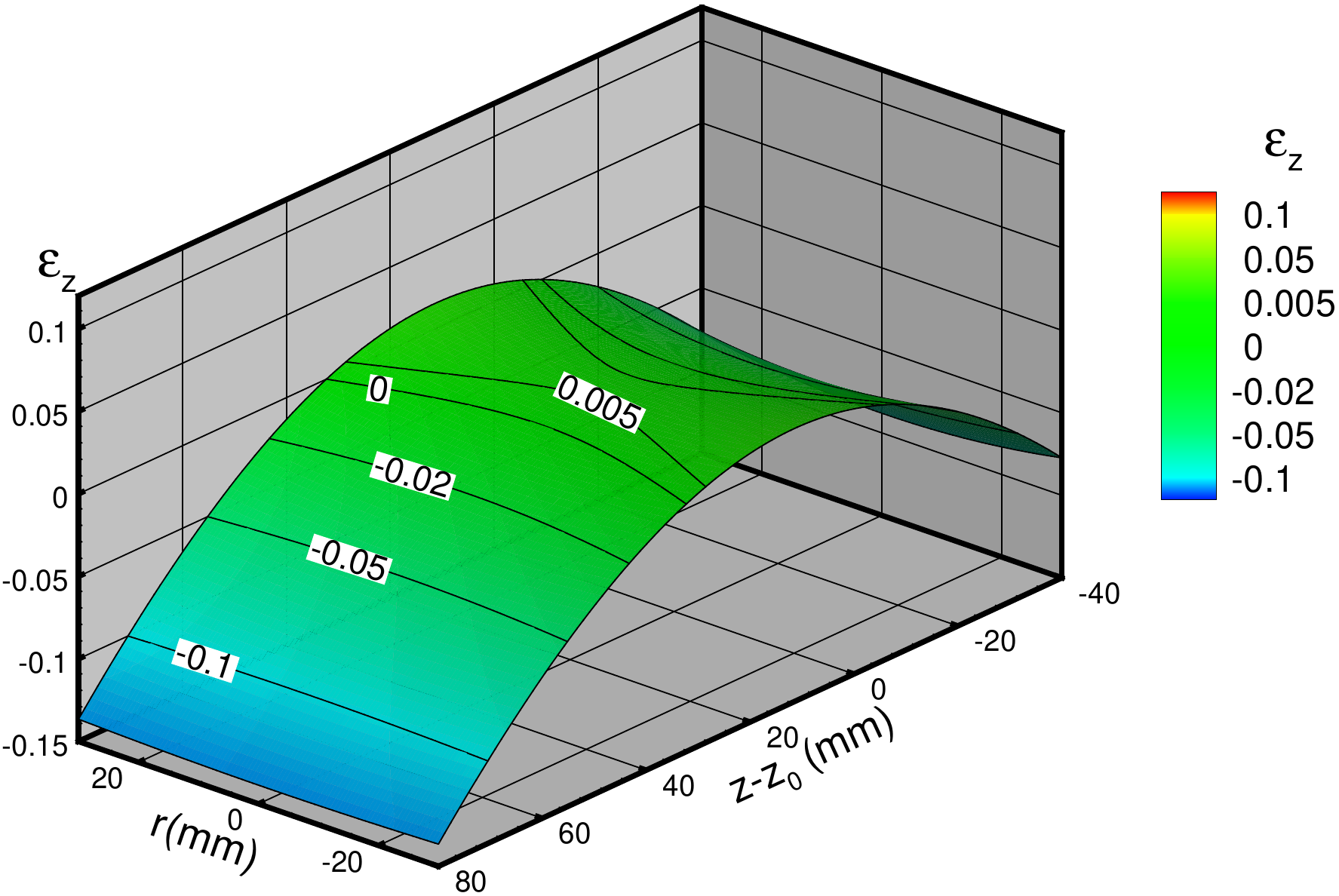}\\(a)\\
\includegraphics[width=\columnwidth, clip]{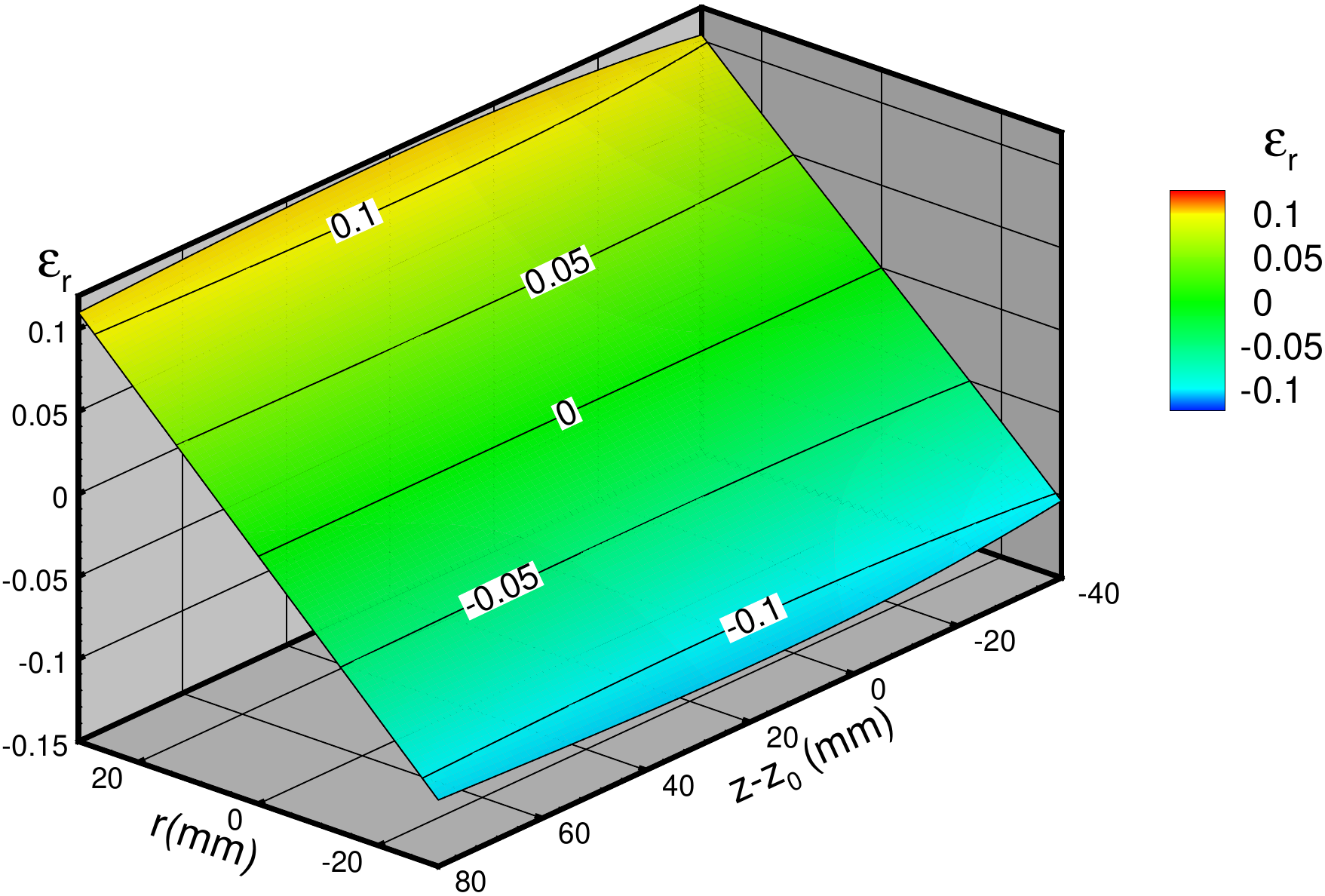}\\(b)
\caption{Magnetic force heterogeneity axial (a) and radial (b) components for the solenoid (height: 555 mm)
of the OLGA installation \citep{MST09}. $r=0$, $z=z_0=-248$~mm is one of two compensation points.
Fig.~\ref{HomSol}a corresponds to the zoomed lower portion of the complete $\nabla (B^2)_z$ curve for the
OLGA solenoid (similar to that of Fig.~\ref{Sol}).} \label{HomSol}
\end{figure}

The spatial variation of the field heterogeneity $\vec{\varepsilon}$ can be calculated numerically when the
magnetic field is known with certainty. This is the case of the magnetic field of a solenoid
(Fig.~\ref{Sol}). In Fig.~\ref{HomSol}, one can locate two compensation points at the axis ($r=0$) at which
$\vec{\varepsilon}=0$. One of them corresponds to the stable  levitation of a bubble; another is unstable
\citep{MST09}.

Because of the cylindrical symmetry, the vector $\vec{\varepsilon}$ has only two components: axial
$\varepsilon_z$ and radial $\varepsilon_r$. It is important to know for the estimation purposes which field
value $B$ is necessary to obtain a given $\vec{\varepsilon}$ inside a sphere of the radius $R$ for a
substance requiring the $|\nabla ({B}^2)|$ value $G$ (see Eq. \ref{comp} and Fig. \ref{gradB2}). The answer
\citep{Quettier} is given by the expression
\begin{equation}\label{mag}
    B=\frac{1}{2}\sqrt{\frac{3GR}{2\varepsilon_r+\varepsilon_z}}.
\end{equation}
In spite of its simplicity, it gives quite accurate results. Two examples can be given for compensation in
water, a case particularly important for life sciences applications. To obtain the gravity heterogeneity
$\varepsilon_z=\varepsilon_r=1$\% inside a sphere of $2R=50$~mm diameter, the magnetic installation should
create, according to (\ref{mag}), the field $B=41$~T. This is close to the world field record obtained with
the hybrid (superconductive+resistive) installations. Such an installation would be extremely expensive. For
$B=16.5$~T, Eq. (\ref{mag}) results in $\varepsilon=1.2$\% for $2R=10$~mm, which corresponds to the existing
installations (Table \ref{tabI}).

\begin{figure}
\centering
\includegraphics[width=0.5\columnwidth]{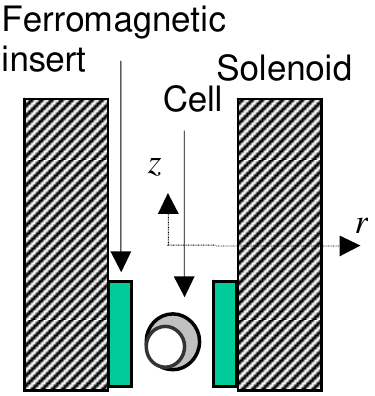}\\
(a)\\
\includegraphics[width=\columnwidth, clip]{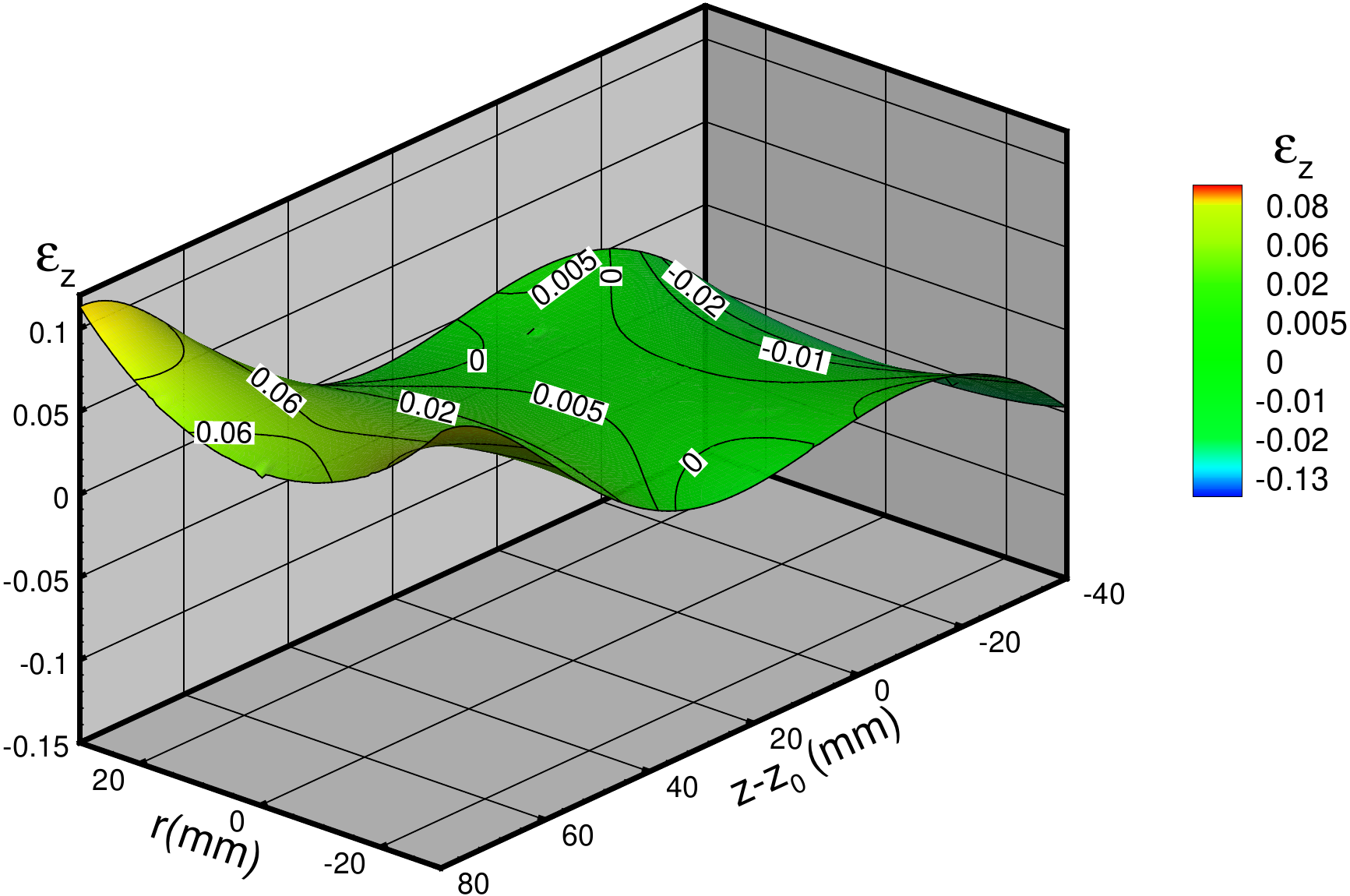}\\
(b)\\\includegraphics[width=\columnwidth, clip]{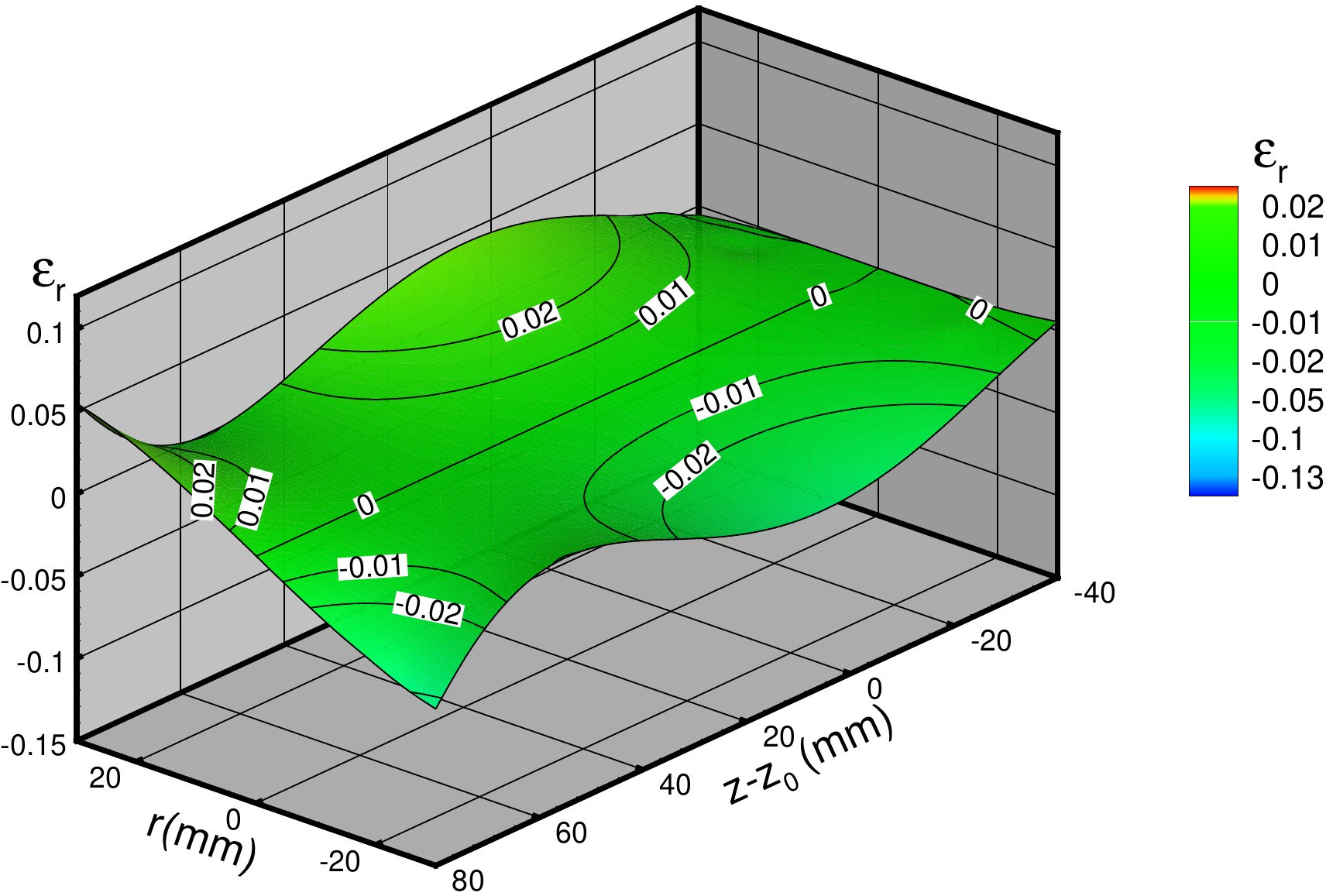}\\
(c) \caption{Insert scheme (a). Axial (b) and radial (c) magnetic force heterogeneities for the same solenoid
as that of Figs. \ref{HomSol} but with the insert. $r=0$, $z=z_0=-155$ mm is the compensation point. The
exact position of the insert with respect to the solenoid center is shown in Fig. \ref{geom-reelle}a below.}
\label{ins}
\end{figure}

Eq. (\ref{mag}) helps finding ways to improve the gravity homogeneity of an existing installation. The local
$B$ increase can be achieved by using ferromagnetic inserts inside the solenoid \citep{Quettier}. It is well
known that the field increases in the vicinity of a ferromagnetic component. The force homogeneity calculated
in presence of the insert  from Fig. \ref{ins}a is shown in Figs. \ref{ins}b,c. The improvement of the radial
heterogeneity is especially large. The calculation of the field has been performed with the Radia freeware
package \citep{Radia} available from the ESRF web site together with its complete description. Comparing to
the case with no insert (Figs. \ref{HomSol}), one obtains an increase of the compensation volume by a factor
5 to 8.

\subsection{Fluid-induced distortion of effective gravity}\label{DistFl}

Let us first consider a two-phase fluid in the \emph{constant} magnetic field and under Earth gravity. Since
$|\chi|\ll 1$ both for liquid and gas phases, a distortion of the background field induced by the liquid and
gas domains and by the interface separating them is usually small. However, it is well known that, in the
electric field, the field distortion can be strongly amplified near the regions of high interface curvature.
Since the equations for the static magnetic field are similar to their electrostatic counterparts (they can
also be expressed in terms of the scalar potential), an analogous effect exists in the magnetic field. The
field distortion is localized in the vicinity of the high curvature interface points. We explain below that
such points can appear in paramagnetic fluids like oxygen.

\begin{figure}
\centering
\includegraphics[width=0.6\columnwidth]{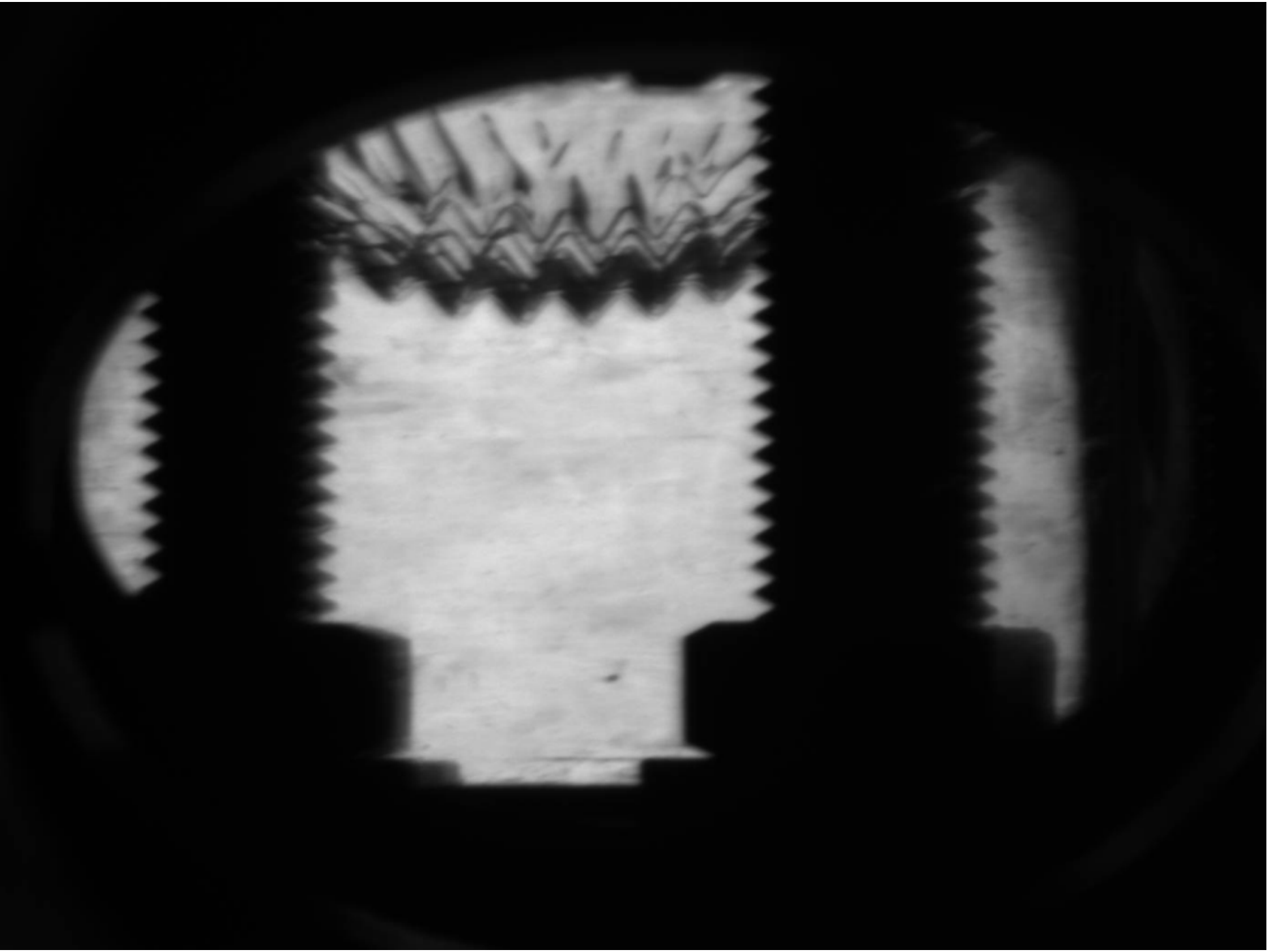}\\
(a)\\
\includegraphics[width=0.6\columnwidth, clip]{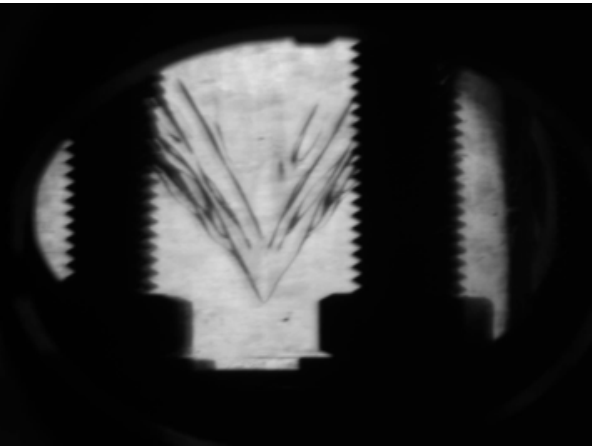}\\
(b) \caption{Surface corrugation (a) and conical bubble shape (b) for oxygen at $T=154.5$K, close to its
critical point ($T_c=154.8$K, $p_c=50$ bar) in OLGA. Two vertical threaded rods that keep the cell together are
visible. Both $B$ and $|\nabla ({B}^2)|$ values for the case (b) are slightly larger than for (a).}
\label{ferro}
\end{figure}

It is well known \citep{Ferromagn67} that the surface of a \emph{ferromagnetic} fluid becomes corrugated when
$B$ exceeds a threshold value $B_c\sim [\sigma g(\rho_L -\rho_V)]^{1/4}$. Here $\sigma$ is the interface
tension; the indices L and V refer to liquid and vapor, respectively. The period of corrugation is
$\lambda=2\pi l_c$, where $l_c=[\sigma/g(\rho_L -\rho_V )]^{1/2}$ is the capillary length. The $\chi$ sign for
paramagnetic substances is the same as for ferromagnetic substances, but the absolute value is much smaller.
For this reason, the same instability occurs for the paramagnetic fluids but at much larger fields, see Fig.
\ref{ferro}a.

When $\lambda/2$ is larger than the cell size, this effect leads to a distortion of the bubble shape. At some
conditions it may develop a conical end (or cusp) \citep{O2sharp2}, see Fig. \ref{ferro}b. This means that
the field is distorted in the vicinity of this end. It is the mutual amplification of the field distortion
and interface deformation that leads to the cusp geometrical singularity \citep{Conical}. This effect is
absent in diamagnetic fluids where an interface deformation induces the field change that causes the
interface smoothing.

When the field with a strong gradient is applied, the situation becomes even more complicated. It is now the
effective gravity acceleration $a_{eff}$ that needs to be used instead of $g$ for the calculation of $B_c$
and $\lambda$. Thus $B_c\to 0$ at compensation conditions \citep{O2inst03} so that the instability always
occurs at compensation. Since $\lambda\to\infty$, the interface corrugation is not observed. According to our
observations, the instability manifests itself by the bubble shape deformation. Far from the critical point,
bubble is of elongated oval shape. Since the instability strength is controlled by the difference $(B-B_c)$,
the elongation should grow with $B$. This leads to an apparent paradox that appears when one uses a
ferromagnetic insert to improve the homogeneity of the background force field (sec. \ref{Solf}). One might
expect an improvement of the bubble sphericity. On the contrary, the bubble deformation grows because the
insert increases $B$ and thus strengthens the instability. Close to the critical point, a cusp appears (Fig.
\ref{ferro}b) because the instability becomes especially strong with the decrease of $B_c\sim [\sigma (\rho_L
-\rho_V)]^{1/4}$.

As mentioned above, this instability is absent in diamagnetic fluids.

\subsection{Container-induced distortion of effective gravity}\label{DistCell}

One needs to be particularly cautious about the materials used for the fabrication of the experimental cell
and its fixation. In practice, the stainless steel is often used because of its strength and high chemical
resistance to corrosion. It is considered to be a non-ferromagnetic material. This is true for a raw piece of
stainless steel. Any mechanical or thermal stress converts at least some superficial layer, adjacent to the
treated surface, to the ferromagnetic state. As an example, one can mention the welding joints. However, the
magnetic strength (i.e. the saturation field) of such components is rather weak. Such a conversion can be
easily demonstrated e.g. with a small but strong rare earth magnet.

\begin{figure}
\centering
\includegraphics[width=0.6\columnwidth]{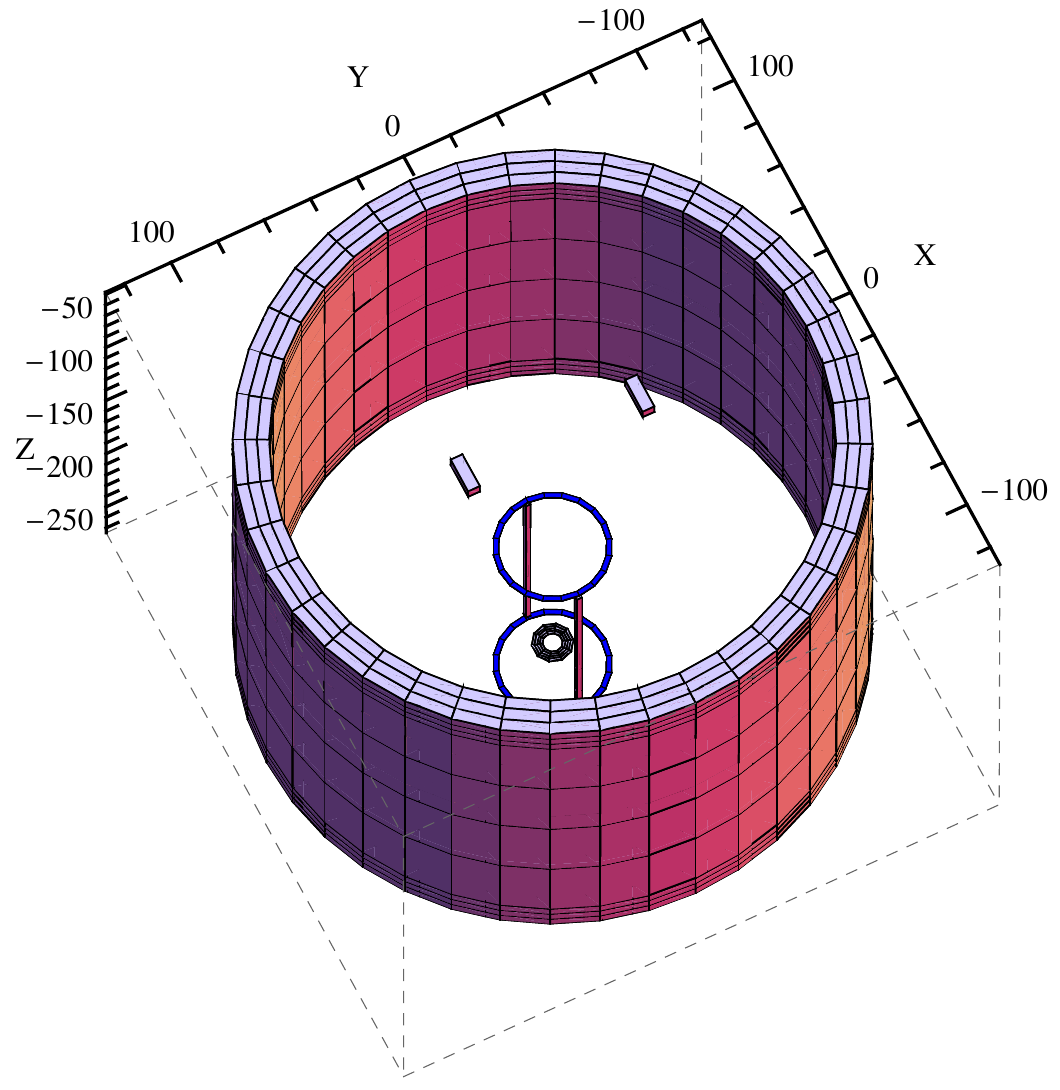}\\
(a)\\
\includegraphics[width=\columnwidth, clip]{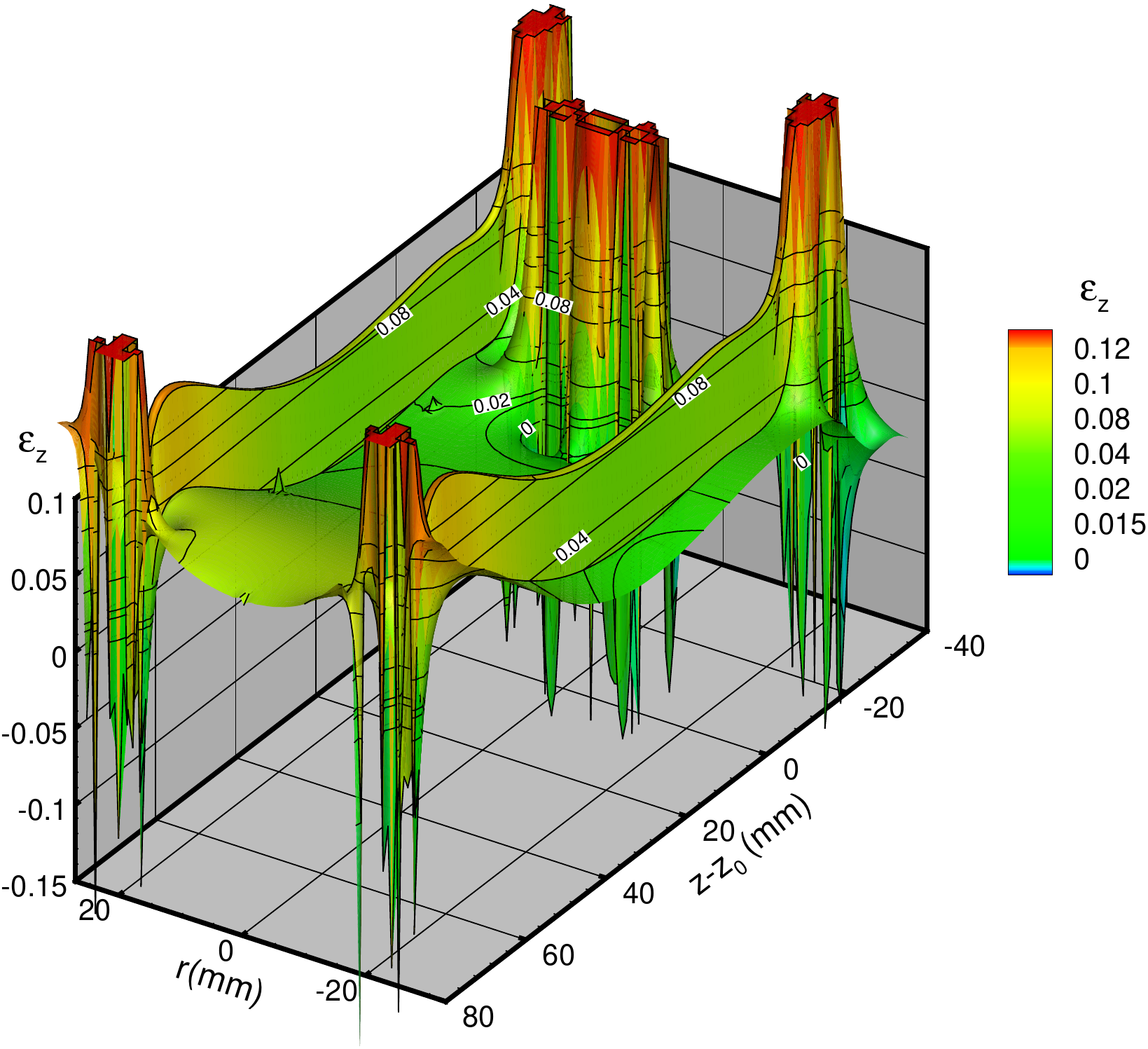}\\
(b) \caption{(a) Several stainless steel cell components inside the OLGA ferromagnetic insert. The position of
the insert with respect to the solenoid center (Fig. \ref{ins}a) is shown with the coordinates in mm. The RADIA
grid used for numerical calculations is visible on the insert. (b) The axial gravity heterogeneity corresponding
to the geometry shown in Fig. \ref{geom-reelle}a, to be compared with Fig. \ref{ins}b.} \label{geom-reelle}
\end{figure}

To estimate the influence of the weakly ferromagnetic structural elements on the effective gravity field,
another field heterogeneity calculation was necessary. Several cell components (Fig. \ref{geom-reelle}a) with
a saturation field of 0.1 T (exaggerated for estimation purposes) has been simulated.
One can see that there is practically no long range force field distortion. It is limited to several mm range
around the element. This means that one can use such elements provided that they are far enough from the
working region. It is better however to avoid the potentially ferromagnetic materials; we replaced stainless
steel by brass or titanium wherever possible.

The influence of one of the simulated elements, a flat stainless steel ring (the smallest of the three rings
shown in Fig. \ref{geom-reelle}a) situated at the bottom of the experimental cell was found experimentally. The
vapor bubble was attracted to the ring if the distance between them was small enough. This attraction
corresponds to the negative $\varepsilon_z$ (Fig. \ref{geom-reelle}b) in the vicinity of the ring, at
$z-z_0\in[0,5]$ mm.

\subsection{Experimental measurement of $\varepsilon$}\label{Exp}

\begin{figure}\sidecaption
\centering
\includegraphics[width=0.3\columnwidth]{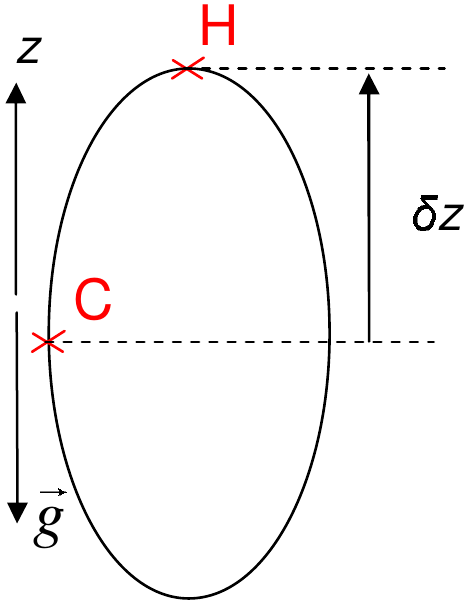}
\caption{Sketch of a vapor bubble magnetically levitated inside the liquid. The bubble is deformed by the
residual effective gravity field.} \label{bubble}
\end{figure}

The experimental measurement of $\vec{\varepsilon}$ is desirable under the compensation conditions. The
background force field testing is possible using a container filled with the liquid and vapor phases of the
same substance. Let container be placed into the magnetic field in such a way that the bubble does not touch
the container walls. Under the ideal (space) weightlessness conditions, the shape of the bubble would be
spherical. Under magnetic compensation, the bubble center is situated at the compensation point or close to
it. Because of the spatial variation of the magnetic force, the bubble becomes elongated (Fig. \ref{bubble}).
From the bubble image, one can measure the bubble surface curvatures $K_H$ and $K_C$ at the points H and C
respectively. The  heterogeneity of the effective gravity acceleration can be estimated (see Appendix A) with the
expression that follows from (\ref{est}),
\begin{equation}\label{eps}
    \varepsilon \approx l_c^2\frac{K_H-K_C}{\delta z}.
\end{equation}
Since $\varepsilon_r$ is neglected in such an estimation, its accuracy is the best for small bubbles. Note that
the sensitivity of this method increases with the decrease of the surface tension. For large $\sigma$, the
bubble remains spherical even at large gravity heterogeneity and it is difficult to measure the difference of
$K_H$ and $K_C$.

More detailed information on the field configuration is obtained if the temperature and pressure of the fluid
can be kept very close to the fluid's liquid-gas critical point (which requires, in general, a precise
thermal regulation). In this case the surface tension can be made extremely small and the corresponding term
can be neglected in (\ref{sh1}). The liquid-vapor interface then follows an equipotential surface for the
"magneto-gravitational" potential \citep{Lorin09},
\begin{equation}\label{U}
    U=\frac{\alpha}{2\mu _0g}B^2-z.
\end{equation}
An equipotential surface (or rather its intersection with the image plane) can thus be visualized directly
\citep{Lorin09}. Different equipotential surfaces are obtained by varying the field or the cell position. The
spatial distribution of the gravity heterogeneity can be found from the shape of the equipotential lines as
$\vec{\varepsilon}=\nabla U$. The latter equation can be established by comparison of (\ref{U}) with
(\ref{aeff}).

Note that the above described methods are not applicable to the case of paramagnetic fluids (sec.
\ref{DistFl}), where the interface deformation and magnetic field distortion are coupled and lead to a strong
self-induced interface deformation even in highly homogeneous effective gravity field.

\section{Concluding remarks}

Magnetic gravity compensation method presents a powerful alternative to the classical low-gravity
experimentation methods involving fluids. It becomes increasingly popular last years, especially for life
sciences applications. About ten installations are available worldwide. The residual gravity heterogeneity
imposes limitations on the applicability of magnetic gravity compensation. The homogeneity of the effective
gravity is related to the magnetic field intensity and can be improved with ferromagnetic inserts. We propose
an original method of measurement of the residual gravity from the bubble shape.

The gravity heterogeneity depends not only on the installation, but also on the sample composition and
structural elements. To reduce the gravity heterogeneity, compounds containing the ferromagnetic substances
(Fe, Co, Ni, etc.) need to be avoided. The heterogeneity is the smallest for single component samples and
needs to be carefully evaluated for multicomponent systems (e.g. for life sciences applications) using the
magnetic susceptibility data for each of the components.

For paramagnetic substances like oxygen, an additional cause of gravity heterogeneity appears because of the
coupling of the magnetic field and deformation of the gas-liquid interfaces. In this case, the effective
residual gravity is more difficult to evaluate because it depends on the interface shape.

\begin{acknowledgements}
The partial financial support by CNES and Air Liquide is gratefully acknowledged.
\end{acknowledgements}

\appendix

\section*{Appendix A: Estimation of the effective gravity acceleration from the shape of a bubble}

The shape of a vapor bubble is determined from an equation that defines a difference of pressures of vapor and
liquid across the interface \citep{HeNoncoal,Cryo02}
\begin{equation}\label{shape}
    \Delta p=K\sigma +\frac{(\chi_L -\chi_V )}{2\mu _0 }B^2-(\rho_L -\rho_V)gz,
\end{equation}
where $K$ is the interface curvature that varies along the interface. Note that the curvature at a given point
of the interface can be measured from its image. At equilibrium, $\Delta p$ is constant along the interface. By
using (\ref{al}), Eq. \ref{shape} becomes
\begin{equation}\label{sh1}
    \Delta p=K\sigma +(\rho_L -\rho_V)\left(\frac{\alpha}{2\mu _0 }B^2-gz\right).
\end{equation}
Let us write this equation for two arbitrary points of the bubble interface and subtract them. By denoting the
difference of a quantity between these points by $\delta$, one obtains, by using (\ref{aeff}),
\begin{equation}
    \sigma\delta K =(\rho_L -\rho_V)\left(g\delta z-\frac{\alpha}{2\mu _0 }\delta(B^2)\right)\approx
    (\rho_L -\rho_V)\,\delta z\; a_{eff,z}.
\end{equation}
Finally, one gets the estimation
\begin{equation}\label{est}
    a_{eff,z} \approx\frac{\delta K}{\delta z}\frac{\sigma }{(\rho _L -\rho _V )}.
\end{equation}

\bibliographystyle{spmpscinat}
\bibliography{MagnLev}   
\end{document}